# Defining best practices in the management of geothermal exploration data

M. DARNET[1], P. CALCAGNO[1], S. HAUKSDOTTIR[2], D. THORBJORNSSON[2], E. TRUMPY[3], J. F. DE WIT[4], T. FRIDRIKSSON[4]

[1]BRGM, France, [2]ISOR, Iceland, [3]CNR, Italy, [4]World Bank, United States

Main contact: m.darnet@brgm.fr

## ABSTRACT

The objective of this work is to define best practices in the management of geothermal exploration data. This study builds on a questionnaire to survey the geothermal data management practices in mature geothermal markets. The inquiry targeted public Regulatory entities with overview of geothermal resources as well as public and private developers. Topics covered in the questionnaire range from the country status to the database set up. The questionnaire focused on the specifications, usage and investments required for installing/maintaining information systems capable of managing exploration data. In addition, information on the different regulatory frameworks and company policies for managing/sharing exploration data has been gathered to identify the requirements imposed on the design of information systems.

The responses were analyzed to identify commonalities in data management practices. They reveal that installing an Information System (IS) is the best practice to systematically and securely manage exploration data. They also provide recommendations with respect to the regulatory framework, data types, data collection methodologies, data storage, data quality control, data accessibility and dissemination, IS architecture, financial investments and human resources required to develop a state-of-the art IS. These results will guide the design of future technical assistance programs for beneficiaries of World Bank support to geothermal exploration activities and it is our belief that they will be beneficial for the geothermal sector at large.

## 1. INTRODUCTION

The World Bank Energy Sector Management Assistance Program’s (ESMAP) Global Geothermal Development Plan (GGDP) aims to scale up geothermal development by mobilizing funding for activities that (i) reduce upstream resource risk, specifically exploration drilling, and (ii) promote dissemination of knowledge and best practices. Proper management of the data collected during exploration is key to secure the sustainability of investments in exploration and thus a priority of the GGDP.

The key objective of the study is to carry out analytical work to identify and define best practices in geothermal exploration data management to serve as a basis for a systematic and secure management of exploration data and hence enhance the sustainability and impact of the investments made within the scope of the GGDP projects. Focus is on geo-scientific data collected during the **exploration** of low and high enthalpy geothermal resources (excluding ground heat pumps) for different geothermal applications (power and heat production). It therefore focuses on the following types of data: surface geological data, surface geophysical data, surface geochemical data, borehole data, well log data, well test data, models based on geothermal exploration data

Target audience is geothermal **Regulators**, e.g. Ministries, Regulatory Institutes, Public Research funding bodies, Research and development institutions, Geological survey organizations and geothermal **Developers**, i.e. companies involved in Geothermal utilization, Geothermal exploration, Geothermal energy distribution, Research & Innovation, Project development…

To this end, we surveyed and synthesized data management practices in the following geothermal markets: France, Germany, Iceland, Indonesia, Italy, Japan, Kenya, Mexico, Netherlands, New Zealand, Philippines and United States. The survey was implemented by two questionnaires, for developers and regulatory entities active in the aforementioned geothermal markets regarding data management. A questionnaire containing 122 questions was designed, put online and shared to developers and a comparable questionnaire containing 149 questions was shared to regulatory institutes in the same countries.

This paper presents a summary of responses to the questionnaires and resulting suggestions for definition of industry best practices in geothermal exploration data management. The suggestions include an estimated cost of installation and maintenance of data management systems in accord with the industry best practices.

## 2. ANALYSIS OF RESPONSES TO QUESTIONNAIRES ON GEOTHERMAL EXPLORATION DATA MANAGEMENT SYSTEMS

In order to define best practices in geothermal data management, the following aspects of the data management practices in mature geothermal markets were studied from both the Developers and Regulators perspectives:

- The types of data collected by developers during the exploration of geothermal resources, including surface exploration and drilling and testing of exploration wells
- The methodology of data collection and approaches to ensure data quality
- What data is shared with public Regulator entities and what subset of this data is made available to the public
- The processes used to disseminate data (including time bound confidentiality, data format, dissemination platforms etc.)

- The software and hardware needs for secure storage of the different types of data and efficient data use (for interpretation, visualization, calculations etc.)
- The investment level required to manage, maintain and use an industry standard geothermal exploration data management system

The section provides a summary of both Regulator and Developer responses respectively, on the aspects considered in recommendations to best practices for data management put forward in chapter 3.

## 2.1 Regulators

Through the questionnaire survey, relevant input from the following geothermal markets was obtained (in alphabetical order): France, Iceland, Italy, The Netherlands, New Zeeland, United States.

Regulatory framework

A regulatory framework for geothermal exploration is in place in the surveyed countries. In general, this framework requires developers to report geothermal exploration data to the regulatory institution. Developers have to comply with time constraints for delivering the data. A standardized reporting workflow is in place for about two thirds of the institutions. The types of data to be reported include surface geology, geophysics and geochemistry, well data, logging and testing, and models computed from these data.

Only half of the regulatory institutes claim to use a standardized workflow when making exploration data accessible to the public. Half of the regulatory institutions explicitly answered that they allow public access to the data, either through web services or upon manual requests. They proceed with a public release generally after an embargo period that depends on the type of data and their sensitivity.

The degree of compliance of geothermal developers with submission of exploration data to the regulatory framework is higher than 50% for two of the surveyed entities, and higher than 80% for three others. The submission is mostly enforced by regular engagements including audits and penalties. Interestingly, one of three institutions from the same country claimed that data submission is not mandatory while the two others answered positively. Although data submission to an official entity is required in all surveyed geothermal markets, it appears that the requirements are not always clear for the parties involved.

Data types and collection

***Geological surface exploration data***

All Regulators said they were collecting the following geological data: structural data, lithological data, hydrogeological data, data on thermal manifestations, geological maps, cross sections and remote sensing data. All the entities manage the geological data within the institution, except one entity that manages remote sensing data through a third party (*Table 1*Table **1**). Only one of the five entities keeps raw data.

Regulators generally use a format for geological data defined by national or international standards. This is interesting in the light of the fact that almost half of the 13 developers that responded to a similar question said that the geological data they submit to Regulators are not structured according to any specific standard.

Majority of Regulators claim that submission of geological data is mandatory. This is in line with the results of similar question answered by developers. All Regulators responding about access of data have geological data either public or accessible through a time-bound access. The data is accessible through a web service or manual request. The frequency of data submission is monthly or annually for Regulators, although most developers said data submission was ad hoc.

| Data | Support | Storage | Type | Standard | Mandatory submission | Confidentiality | Data Management | Access | Frequency |
|---|---|---|---|---|---|---|---|---|---|
| Geological data | ☐ Paper<br>☐ Tape<br>☐ CD-DVD<br>☐ Hard disk<br>☒ Other: | ☐ Warehouse<br>☐ Local Directory<br>☒ Directory on server<br>☐ Local database<br>☒ Database on server | ☐ Raw<br>☐ Processed<br>☒ Interpreted | ☐ None<br>☒ National<br>☒ International<br>☐ Industry | ☒ Yes<br>☐ No | ☒ Public<br>☐ Confidential<br>☒ Time-bound | ☒ In house<br>☐ Subcontracted<br>☐ Supplied from third party | ☐ Not allowed<br>☒ Manual request<br>☒ Web service<br>☐ Query script | ☐ Annually or less frequently<br>☒ Monthly to annually<br>☐ More frequently than monthly<br>☐ Ad hoc |

Table 1*: Summary of responses by Regulators regarding geological surface exploration data.*

***Geophysical surface exploration data***

Regulators collect and/or manage the following type of geophysical data: gravimetric surveys, geomagnetic surveys, MT/CSEM surveys, ERT surveys, self-potential surveys, passive seismic surveys, active seismic surveys and ground temperature/heat flow mapping. As presented in the summary (*Table 2*), geophysical data is almost in all cases managed in-house, the format being hard-disc, on paper or other unspecified data format. Regulators seem to use various methods for standards for geophysical data whereas developers use mainly international as well as industrial standards.

Mandatory submission is required by most Regulators of all geophysical data listed in the questions. These answers are interesting in the light of the answers to a similar question from developers. According to answers from developers, the mandatory submission of is geophysical data is generally not required.

All Regulators replying to the survey give access to geophysical surface exploration data, either public access or time-bound access. but the opposite was seen in answers from developers, i.e. in about 80% cases, geophysical data is handled as confidential data.

Annual or less frequent submission is the most common practice, although both more frequent and ad hoc submission is in some cases required. According to developers, ad hoc data submission is the most common practice.

| Data | Support | Storage | Type | Standard | Mandatory submission | Confidentiality | Data Management | Access | Frequency |
|---|---|---|---|---|---|---|---|---|---|
| Geophysical data | ☒ Paper<br>☐ Tape<br>☐ CD-DVD<br>☒ Hard disk<br>☒ Other: | ☐ Warehouse<br>☐ Local Directory<br>☒ Directory on server<br>☐ Local database<br>☒ Database on server | ☐ Raw<br>☐ Processed<br>☒ Interpreted | ☒ None<br>☒ National<br>☒ International<br>☒ Industry | ☒ Yes<br>☐ No | ☒ Public<br>☐ Confidential<br>☒ Time-bound | ☒ In house<br>☐ Subcontracted<br>☐ Supplied from third party | ☐ Not allowed<br>☒ Manual request<br>☒ Web service<br>☐ Query script | ☒ Annually or less frequently<br>☒ Monthly to annually<br>☐ More frequently than monthly<br>☐ Ad hoc |

*Table 2 : Summary of responses by Regulators regarding geophysical surface exploration data.*

***Geochemical surface exploration data***

Regulators generally collect the following types of geochemical data: active geothermal features, fluid/gas composition, fluid/gas samples, chemical geothermometers and soil gas/flux analysis. *Table 3* shows a summary of the most common answers from the five regulatory entities that gave answers.

Regulators generally manage data in-house, on paper, hard-disc or more commonly in other undefined format. All Regulators except one, claim that submission of geochemical data submission is mandatory but developers state that data submission is not mandatory for any of the data types listed in the question.

Access to the data is either public- or time-bound access and in about 75% cases Regulators allow access based on manual request and/or web service. This is different from the developers who in most cases define their geochemical data as confidential data.

Monthly to annual data submission is the most common frequency of data submission, but according to answers from developers, more frequent than monthly or ad hoc data submission appears to be the most common practise.

| Data | Support | Storage | Type | Standard | Mandatory submission | Confidentiality | Data Management | Access | Frequency |
|---|---|---|---|---|---|---|---|---|---|
| Geochemical data | ☒ Paper<br>☐ Tape<br>☐ CD-DVD<br>☒ Hard disk<br>☒ Other: | ☐ Warehouse<br>☐ Local Directory<br>☒ Directory on server<br>☐ Local database<br>☒ Database on server | ☐ Raw<br>☐ Processed<br>☒ Interpreted | ☒ None<br>☒ National<br>☒ International<br>☒ Industry | ☒ Yes<br>☐ No | ☒ Public<br>☐ Confidential<br>☒ Time-bound | ☒ In house<br>☐ Subcontracted<br>☐ Supplied from third party | ☐ Not allowed<br>☒ Manual request<br>☒ Web service<br>☐ Query script | ☐ Annually or less frequently<br>☒ Monthly to annually<br>☐ More frequently than monthly<br>☐ Ad hoc |

*Table 3 : Summary of responses by Regulators regarding geochemical surface exploration data.*

**Borehole data**

Several different data types were listed in the questionnaire; well design (trajectory, diameter, casing scheme), drilling parameters (ROP, mud weight, etc.), Drilling reports (daily/well completion reports), end of well reports (location, type, owner, date, rig, etc.), samples of cores/cuttings and description of cores/cuttings (petrography, lithology, alteration, mineralogy).

Generally Regulators manage borehole data in-house with the exceptions of cores/cuttings that are also managed by third party (*Table 4***Error! Reference source not found.**). More than 50% of the Regulators store their borehole data on other undefined format whereas all the developers store most of their borehole data on hard-discs.

Apparently, most Regulators store borehole data in directories or databases on servers, and claim to structure borehole data in a specific data format defined by all the listed standards, i.e. national standards, international standards, industry standards and other undefined standards.

Data submission is mandatory for all the borehole data types listed in the relevant question with the exception that one Regulator stated that delivery of core/cutting samples was not mandatory. This is interesting as most developers stated that submission of borehole data was not mandatory. All the Regulators allow access to borehole data, either public or time-bound, most upon manual request, but some allow access through web services.

| Data | Support | Storage | Type | Standard | Mandatory submission | Confidentiality | Data Management | Access | Frequency |
|---|---|---|---|---|---|---|---|---|---|
| Borehole data | ☒ Paper<br>☐ Tape<br>☐ CD-DVD<br>☒ Hard disk<br>☒ Other: | ☐ Warehouse<br>☒ Local Directory<br>☒ Directory on server<br>☐ Local database<br>☒ Database on server | ☐ Raw<br>☒ Processed<br>☒ Interpreted | ☐ None<br>☒ National<br>☒ International<br>☐ Industry | ☒ Yes<br>☐ No | ☒ Public<br>☐ Confidential<br>☒ Time-bound | ☒ In house<br>☐ Subcontracted<br>☒ Supplied from third party | ☐ Not allowed<br>☒ Manual request<br>☒ Web service<br>☐ Query script | ☒ Annually or less frequently<br>☒ Monthly to annually<br>☒ More frequently than monthly<br>☒ Ad hoc |

*Table 4 : Summary of responses by Regulators regarding borehole exploration data.*

**Well logging data**

Regulators generally collect the following types of borehole data: logging while drilling (gamma ray, resistivity, etc.), caliper logs, borehole geophysics, cement bound logs, temperature and pressure logs, resistivity, porosity (density, neutron porosity, sonic logs), lithology logs (gamma ray, self-potential), fluid retrieved analysis (chemical and physical), downhole video, televiewer, nuclear magnetic resonance logs. Generally, Regulators manage well logging data and collect data on retrieved fluid whereas fewer collect downhole videos, televiewer data and nuclear magnetic logs. All the Regulators collect and/or manage the other listed data types. Well logging data is equally often managed in-house or supplied from third party (*Table 5*).

In most cases, Regulators store their well logging data in an undefined format whereas developers most commonly use hard-discs and paper format. The type of well logging data is in most cases (depending on the data type) considered to be interpreted data but less as processed/quality controlled or processed in other undefined way. This is in line with answers from developers.

Commonly Regulators structure well logging data according to industry standards or use either international or other undefined standards. While Regulators claim that submission of all the listed well logging data types is mandatory the answers from developers suggest the opposite. Unlike the Regulators, most developers define their well logging data as confidential data.

| Data | Support | Storage | Type | Standard | Mandatory submission | Confidentiality | Data Management | Access | Frequency |
|---|---|---|---|---|---|---|---|---|---|
| **Well logging data** | ☒ Paper<br>☐ Tape<br>☐ CD-DVD<br>☒ Hard disk<br>☒ Other: | ☒ Warehouse<br>☒ Local Directory<br>☒ Directory on server<br>☒ Local database<br>☒ Database on server | ☐ Raw<br>☒ Processed<br>☒ Interpreted | ☐ None<br>☐ National<br>☒ International<br>☒ Industry | ☒ Yes<br>☐ No | ☒ Public<br>☐ Confidential<br>☒ Time-bound | ☒ In house<br>☐ Subcontracted<br>☒ Supplied from third party | ☐ Not allowed<br>☒ Manual request<br>☒ Web service<br>☐ Query script | ☒ Annually or less frequently<br>☒ Monthly to annually<br>☒ More frequently than monthly<br>☒ Ad hoc |

*Table 5 : Summary of responses by Regulators regarding well logging exploration data.*

**Well testing data**

Regulators generally collect the following type of well testing data: fluid temperature, fluid pressure, flow rate (including spinner logs), well discharge, tracer flow testing and well test report. The answers reveal that Regulators and developers collect and/or manage all these types of data.

Four of five Regulators manage the data in-house or rely on a third party but one (*Table 6*) manages the data using a subcontractor. In most cases Regulators use an undefined format for the well testing data, but two out of five entities use paper format.

In most cases, Regulators state that their well testing data is either raw data or interpreted.

Data managed by Regulators is defined mostly by industrial standards. The responses on access rights, suggest that Regulators either allow public or time-bound access to well testing data, either upon manual request or through a web service.

| Data | Support | Storage | Type | Standard | Mandatory submission | Confidentiality | Data Management | Access | Frequency |
|---|---|---|---|---|---|---|---|---|---|
| **Well testing data** | ☒ Paper<br>☐ Tape<br>☐ CD-DVD<br>☒ Hard disk<br>☒ Other: | ☒ Warehouse<br>☐ Local Directory<br>☒ Directory on server<br>☒ Local database<br>☒ Database on server | ☒ Raw<br>☐ Processed<br>☒ Interpreted | ☒ None<br>☐ National<br>☒ International<br>☒ Industry | ☐ Yes<br>☐ No | ☒ Public<br>☐ Confidential<br>☒ Time-bound | ☒ In house<br>☒ Subcontracted<br>☒ Supplied from third party | ☐ Not allowed<br>☒ Manual request<br>☒ Web service<br>☐ Query script | ☒ Annually or less frequently<br>☒ Monthly to annually<br>☒ More frequently than monthly<br>☒ Ad hoc |

*Table 6 : Summary of responses by Regulators regarding well testing exploration data.*

**Models**

In the questionnaire sent to Regulators several different model types were listed: conceptual model, numerical static geological model, geophysical model, geochemical model, hydrogeological model, thermal model, geomechanical model, coupled hydrogeological and/or thermal and/or geomechanical model and geothermal resource assessment. All Regulators who responded collect and/or manage all of the data listed except one entity that does not manage geomechanical, coupled and geothermal resource assessment models. Half of the Regulators manage their models in-house and the other half either relies on supply from third party or other undefined method (

| Data | Software | Storage | Mandatory submission | Confidentiality | Data Management | Access | Frequency |
|---|---|---|---|---|---|---|---|
| **Models** | ☒ Proprietary<br>☒ Commercial<br>☒ Open source<br>☒ Other: | ☒ Local Directory<br>☒ Directory on server<br>☐ Local database<br>☒ Database on server<br>☒ Open source | ☐ Yes<br>☒ No | ☒ Public<br>☐ Confidential<br>☒ Time-bound | ☒ In house<br>☐ Subcontracted<br>☒ Supplied from third party | ☐ Not allowed<br>☒ Manual request<br>☒ Web service<br>☐ Query script | ☐ Annually or less frequently<br>☒ Monthly to annually<br>☐ More frequently than monthly<br>☒ Ad hoc |

*Table 7*).

| Data | Software | Storage | Mandatory submission | Confidentiality | Data Management | Access | Frequency |
|---|---|---|---|---|---|---|---|
| **Models** | ☒ Proprietary<br>☒ Commercial<br>☒ Open source<br>☒ Other: | ☒ Local Directory<br>☒ Directory on server<br>☐ Local database<br>☒ Database on server<br>☒ Open source | ☐ Yes<br>☒ No | ☒ Public<br>☐ Confidential<br>☒ Time-bound | ☒ In house<br>☐ Subcontracted<br>☒ Supplied from third party | ☐ Not allowed<br>☒ Manual request<br>☒ Web service<br>☐ Query script | ☐ Annually or less frequently<br>☒ Monthly to annually<br>☐ More frequently than monthly<br>☒ Ad hoc |

*Table 7 : Summary of responses by Regulators regarding models.*

System architecture, storage, quality control and ownership

Most Regulators have an IS in place. It can be interpreted that entities that consider themselves as not having an IS either have their data managed or coordinated by a third party (25%) or that there do not have any system in use for their data.

The main IS architecture developed by Regulators in mature geothermal markets is based on a central system that produces new/restructures information but one entity has developed a simple feature/index approach. The IS developed are mostly based on relational database (2/3 of the cases) but unstructured database are also used (1/3 of the cases).

Apparently, there are no preferred database management systems for geothermal exploration data. Several different types are used by the Regulators, e.g. ESRI – Geodatabase, Microsoft Access, Microsoft SQL server, Oracle, PostgreSQL, MongoDB, but the number of answers is not enough to see any preferences they might have in general. The majority of data for regulatory entities is hosted locally using relational databases (80% of the cases).

From the all entities having an IS in place, only a small fraction have a defined or verified quality control system. This is also true for the use of other standards (e.g. semantical standards). The majority of entities makes use of discipline experts to control the quality of data.

Regulators generally have data readily available (24/7). Safeguarding protocols to secure the storage are in place for all of the Regulators, including the use of both firewall and system backup. Ownership of the hardware related to the IS is equally public and private for the Regulators, but less than half of the entities respond to this question.

Accessibility and dissemination

Statistics on data access constraints to the public are:

- *Visualizing data*: 66% Unrestricted, 33% Restricted
- *Obtaining data through an order*: 50% Free unrestricted, 25% Paid for service unrestricted, 25% Free restricted
- *Online downloading of public data*: 50% Yes, 50% Yes, in the future

The type of platform or software used to disseminate data are: Download services, Web Services, Web feature Services, Web Map Service, Portals.

To visualize and share data, most entities use “Light” software (only able to share and visualize data). This software is either developed in house or belongs to a third party or Open Source. It must be noted that only one entity is using a “Heavy” software (able to share and visualize data but also to process data).

Cost of Data Information System Installation

Responses from Regulators indicate that developing and implementing IS for exploration data can be expected to take 1 to 5 years and in general such systems are planned to be operational for more than 20 years**.** The estimated initial investment (including salaries, cost for hardware and software) is $100,000$ - $1,000,000 and 26-50 person-months workload were involved in the development of the IS for managing geothermal exploration data (cumulatively since commissioning of the system). The estimated budget for maintenance and operation (including salaries, cost for hardware and software) is $10,000 – $100,000/year. Less than 3 persons/year are involved in the maintenance and operation of the system**.**

When the IS for regulatory entity deals with more than geothermal data (e.g. oil and gas wells, mineral resources), the amounts are larger but have been excluded from our analysis since it falls outside the scope of the assignment.

### 2.2 Developers

Thirteen companies from the following geothermal markets have provided relevant input to our questionnaire survey (in alphabetical order): France, Germany, Iceland, Indonesia, Italy, Kenya and New Zeeland.

Data types and collection

Seven out of thirteen developers who answered the survey collect data systematically according to their own definition of systematic data collection and four out of the seven remaining are planning to implement an IS for managing geothermal exploration data. A bigger portion of the companies (ten entities) is collaborating with other companies in data collection whereas five and six developers are cooperating in data storage and data management, respectively.

**Geological surface exploration data**

Developers collect and/or manage the following types of geological data: structural data, lithological data, hydrogeological data, data on thermal manifestations, geological maps, geological cross sections and remote sensing data. Most developers manage their surface geological data in-house whereas the management is subcontracted in 8-15% cases, depending on data type (*Table 8*).

Most geological data managed by developers is either processed (50%) or interpreted (33%). This was observed for all data types except remote sensing where interpreted data was more common. Almost half of the developers do not structure the data according to standards. Only two developers state using international standards or industry standards for most data types.

According to answers from developers, mandatory submission of geological surface data is generally not required (60-85%). Most data collected and managed by developers is considered confidential. Two developers either allow public access or time bound access to the geological data but most developers (about 70%) allow third parties access to their data through a manual request.

Developers are generally required to submit data monthly to annually or ad hoc.

| Data | Support | Storage | Type | Standard | Mandatory submission | Confidentiality | Data Management | Access | Frequency |
|---|---|---|---|---|---|---|---|---|---|
| Geological data | ☒ Paper<br>☐ Tape<br>☐ CD-DVD<br>☒ Hard disk<br>☐ Other: | ☐ Warehouse<br>☐ Local Directory<br>☒ Directory on server<br>☒ Local database<br>☒ Database on server | ☐ Raw<br>☒ Processed<br>☒ Interpreted | ☒ None<br>☐ National<br>☐ International<br>☐ Industry | ☐ Yes<br>☒ No | ☒ Public<br>☒ Confidential<br>☐ Time-bound | ☒ In house<br>☐ Subcontracted<br>☒ Supplied from third party | ☒ Not allowed<br>☒ Manual request<br>☐ Web service<br>☐ Query script | ☐ Annually or less frequently<br>☒ Monthly to annually<br>☐ More frequently than monthly<br>☒ Ad hoc |

*Table 8 : Summary of responses by developers regarding geological exploration data.*

**Geophysical surface exploration data**

Several different data types were listed in the survey focusing on surface geophysical data: gravimetric surveys, geomagnetic surveys, MT/CSEM surveys, ERT surveys, self-potential surveys, passive seismic surveys, active seismic surveys and ground temperature/heat flow mapping. Most developers (92%) who responded collect and/or manage MT/CSEM and passive seismic data and about 70-77% collect and/or manage gravimetric data, geomagnetic data and ground temperature data. Data from electrical resistivity surveys (ERT) and active seismic surveys are collected and/or managed by about 62% of the developers.

In most cases (60-70%), geophysical surface exploration data is managed in-house (*Table 9*). In other cases (12-25%), data management is either subcontracted or supplied from third party. Geophysical data is stored on hard-disc in about 90% cases except for data from self-potential surveys that are stored on hard-disc in about 60% cases. Although the most common practise to store geophysical data is in directories on servers, other practises are also common (e.g. local directory, directory on server, database on server). In most cases, geophysical data has been interpreted. When the geophysical data has not been interpreted, it has in most cases either been processed or is quality controlled in some way. According to the survey, developers use international standards or no specific format for their geophysical data but industry standards are also common.

Answers regarding mandatory submission of geophysical data reveal that developers and Regulators are not in agreement. According to developers, no mandatory submission of data is required, except in one and two cases for seismic and ground temperature data. The opposite is observed in the answers from Regulators stating that mandatory submission is required for all geophysical data types.

About 80% of developers handle their geophysical data as confidential. On the other hand, all the Regulators who answered similar question give access to geophysical data, either time-bound access or the data is handled as public data. However, most developers state that their geophysical data can be accessed upon manual request. In most cases, Regulators also give access to geophysical data upon manual request.

Ad hoc submission of geophysical data appears to be the most common practice although annual or less frequent submission is also common which is in agreement to answers from Regulators.

| Data | Support | Storage | Type | Standard | Mandatory submission | Confidentiality | Data Management | Access | Frequency |
|---|---|---|---|---|---|---|---|---|---|
| Geophysical data | ☒ Paper<br>☐ Tape<br>☐ CD-DVD<br>☒ Hard disk<br>☒ Other: | ☒ Warehouse<br>☒ Local Directory<br>☒ Directory on server<br>☒ Local database<br>☒ Database on server | ☐ Raw<br>☒ Processed<br>☒ Interpreted | ☒ None<br>☐ National<br>☒ International<br>☒ Industry | ☐ Yes<br>☒ No | ☒ Public<br>☒ Confidential<br>☐ Time-bound | ☒ In house<br>☐ Subcontracted<br>☒ Supplied from third party | ☒ Not allowed<br>☒ Manual request<br>☒ Web service<br>☐ Query script | ☒ Annually or less frequently<br>☒ Monthly to annually<br>☐ More frequently than monthly<br>☒ Ad hoc |

*Table 9 : Summary of responses by developers regarding geophysical exploration data.*

**Geochemical surface exploration data**

Between 80-90% of the developers who responded to question on the type of geochemical data collected, managed or maintained within their company collect all the data types listed in the question (active geothermal features, fluid/gas composition, fluid/gas samples, chemical geothermometers and soil gas/flux analysis). Most developers (66-70%) manage their data in-house, whereas 10-20% either rely on subcontractors or supply from third party (*Table 10*). In about 70-90% cases, developers are storing their data on hard-disks and in about 50% cases, data is also stored on paper.

Data on active geothermal features, fluid/gas samples and geothermometers are most commonly stored in databases on servers (about 60-70%) in each case, whereas other data is more distributed between different storages (e.g. local directory, directory on server, local database).

Developers were given the option to choose between *raw data, processed data, quality-controlled data, interpreted data* and *other* on types of data processing. In most cases, data have been interpreted (40-55%). Most developers (50-55%) do not structure their data in a specific format as defined by standards. About 10-20% of the developers structure their geochemical data as defined by national standards, international standards or industry standards.

Only one out of ten developers said that submission of data on active geothermal features and fluid/gas composition was mandatory. In all other cases for all the different geochemical data types, data submission is not mandatory according to the responses of the developers. The answers by Regulators to the same question show the opposite results, claiming that submission of all data types is mandatory. Only one developer allows public access to its geochemical data whereas all the other who responded to a question on data access handle their data as confidential.

| Data | Support | Storage | Type | Standard | Mandatory submission | Confidentiality | Data Management | Access | Frequency |
|---|---|---|---|---|---|---|---|---|---|
| **Geochemical data** | ☒ Paper<br>☐ Tape<br>☐ CD-DVD<br>☒ Hard disk<br>☐ Other: | ☒ Warehouse<br>☒ Local Directory<br>☒ Directory on server<br>☒ Local database<br>☒ Database on server | ☐ Raw<br>☒ Processed<br>☒ Quality controlled<br>☒ Interpreted | ☒ None<br>☐ National<br>☐ International<br>☐ Industry | ☐ Yes<br>☒ No | ☐ Public<br>☒ Confidential<br>☐ Time-bound | ☒ In house<br>☐ Subcontracted<br>☐ Supplied from third party | ☒ Not allowed<br>☒ Manual request<br>☐ Web service<br>☐ Query script | ☒ Annually or less frequently<br>☒ Monthly to annually<br>☒ More frequently than monthly<br>☒ Ad hoc |

*Table 10 : Summary of responses by developers regarding geochemical exploration data.*

**Borehole data**

Developers generally deal with the following types of wellbore data: well design (trajectory, diameter, casing scheme), drilling parameters (ROP, mud weight, etc.), drilling reports (daily/well completion reports), end of well reports (location, type, owner, date, rig, etc.), samples of cores/cuttings and description of cores/cuttings (petrography, lithology, alteration, mineralogy). About 75% of the developers manage their borehole data in-house whereas about 25% of the developers rely either on supply from third parties or subcontractors (*Table 11*). Apart from samples of cores/cuttings and end of well reports, all the developers who responded to a question on data format store the borehole data on hard-disc. Most developers, or 67 and 92% store data on cores/cuttings and end of well reports on hard-disc, respectively. About 75% of the developers also store well design data, end of well reports and descriptions of cores/cuttings on paper and about 60% of the developers also store their drilling reports and data on cores/cuttings on paper.

| Data | Support | Storage | Type | Standard | Mandatory submission | Confidentiality | Data Management | Access | Frequency |
|---|---|---|---|---|---|---|---|---|---|
| **Borehole data** | ☒ Paper<br>☐ Tape<br>☐ CD-DVD<br>☒ Hard disk<br>☐ Other: | ☒ Warehouse<br>☒ Local Directory<br>☒ Directory on server<br>☒ Local database<br>☒ Database on server | ☐ Raw<br>☒ Processed<br>☒ Quality controlled<br>☒ Interpreted | ☒ None<br>☐ National<br>☐ International<br>☐ Industry | ☐ Yes<br>☒ No | ☐ Public<br>☒ Confidential<br>☐ Time-bound | ☒ In house<br>☒ Subcontracted<br>☒ Supplied from third party | ☐ Not allowed<br>☒ Manual request<br>☒ Web service<br>☐ Query script | ☒ Annually or less frequently<br>☒ Monthly to annually<br>☒ More frequently than monthly<br>☒ Ad hoc |

*Table 11 : Summary of responses by developers regarding borehole exploration data.*

Storage in directories on servers, databases on servers and local databases are the most common ways to store borehole data. Warehouses and local directories are also used to store borehole data.

From the results of the survey it is evident that due to the nature of borehole data it is rather stored as “quality controlled” rather than “interpreted” or “processed” as is more common for other data types. It is interesting to see that developers regard descriptions of cores/cuttings rightfully to be “interpreted” data. Most developers claim that their borehole data is not structured in a specific format as defined by national or international standards. Some developers use industry standards, most commonly for well designs and end of well reports.

Regulators state that submission of all types of borehole data is mandatory whereas most developers generally stated that submission of borehole data is not mandatory. More than 75% of the developers define borehole data as confidential data. Despite this, about 70% of developers allow access upon manual request and some (8-23%, depending on data type) allow access through web service.

**Well logging data**

The majority of developers collect and/or manage caliper-, cement bound-, temperature-, pressure-, lithology-, and resistivity logs as well as televiewer data. Most developers manage their well logging data in-house (50-70% depending on the data type). In other cases, developers either rely on subcontractors or supply from third parties.

Although the well log data is in most cases on hard-discs, paper format, CD/DVD, other undefined formats are also common. It is interesting that the most common format among the developers is the third most common among Regulators, who in most cases store their well logging data on undefined format. More than 50% of developers store their well logging data either in directories on servers or in databases on servers. In about 20-40% cases, well logging data is stored in local directories.

According to answers to the survey, most developers manage interpreted well logging data (*Table 12*) and this is in line with the results from Regulators. Most developers claim that they use industry standards to structure well logging data. Those who do not apply industry standards do not identify any specific data format. The main difference between Regulators and developers is that the Regulators who do not apply industry standards claim they apply either international or other undefined standards, none claim to use any unspecific format.

Regulators assume submission of all the data types listed in the questions on well logging data, most developers claim that they are not required to submit well logging data. Most developers (80-90%) define their well logging data as confidential data. The Regulators, who usually are receiving submitted data, do not define the data as confidential. Although this is the case, most developers and Regulators allow access to well logging data upon manual request according to the questionnaires.

| Data | Support | Storage | Type | Standard | Mandatory submission | Confidentiality | Data Management | Access | Frequency |
|---|---|---|---|---|---|---|---|---|---|
| **Well logging data** | ☒ Paper<br>☐ Tape<br>☒ CD-DVD<br>☒ Hard disk<br>☒ Other: | ☐ Warehouse<br>☒ Local Directory<br>☒ Directory on server<br>☒ Local database<br>☒ Database on server | ☐ Raw<br>☒ Processed<br>☐ Quality controlled<br>☒ Interpreted | ☒ None<br>☐ National<br>☐ International<br>☒ Industry | ☐ Yes<br>☒ No | ☐ Public<br>☒ Confidential<br>☐ Time-bound | ☒ In house<br>☒ Subcontracted<br>☒ Supplied from third party | ☒ Not allowed<br>☒ Manual request<br>☒ Web service<br>☐ Query script | ☒ Annually or less frequently<br>☐ Monthly to annually<br>☐ More frequently than monthly<br>☒ Ad hoc |

*Table 12 : Summary of responses by* developers *regarding well logging exploration data.*

**Well testing data**

The data types listed for well testing in the survey included; Fluid temperature, fluid pressure, flowrate (including spinner logs), well discharge, tracer flow testing and well test report. About 85 to 100% of the developers who responded on well testing data collect and/or manage the data types listed above and 77% collect and/or manage tracer flow testing data. Similar results are seen when Regulators answer the same question. In most cases (>75%), developers manage their well testing data in-house. About 90% of the developers store their well testing data on hard-discs and among 60-70% of the developers, data is available on paper (*Table 13*). Around one-third of the developers store their well testing data on CD/DVD.

Although the most common storage method is to store well testing data in directories on servers, other methods are also common (e.g. local directory, local database, database on server). In about 80% cases, well testing data has either been interpreted or processed to some extent but all have undergone quality control. These results are a bit different from answers from Regulators who in most cases state that their well testing data is either interpreted or stored as raw data.

In 60-70% cases, developers said they structured their well testing data in format defined by industry standards. In all other cases, developers do not have any specified format for the well testing data. According to almost all developers mandatory submission of well testing data is not required. This is in contradiction with the answers from Regulators who state that mandatory submission of well testing data is required.

In 85-100% cases, developers define their well testing data as confidential data. However, about 75% of the developers allow access to well testing data upon manual request. At the same time, Regulators who require mandatory data submission define their data as public or they allow time-bound access to well testing data.

| Data | Support | Storage | Type | Standard | Mandatory submission | Confidentiality | Data Management | Access | Frequency |
|---|---|---|---|---|---|---|---|---|---|
| **Well testing data** | ☒ Paper<br>☐ Tape<br>☒ CD-DVD<br>☒ Hard disk<br>☐ Other: | ☐ Warehouse<br>☒ Local Directory<br>☒ Directory on server<br>☒ Local database<br>☒ Database on server | ☐ Raw<br>☒ Processed<br>☒ Quality controlled<br>☒ Interpreted | ☒ None<br>☐ National<br>☐ International<br>☒ Industry | ☐ Yes<br>☒ No | ☐ Public<br>☒ Confidential<br>☐ Time-bound | ☒ In house<br>☐ Subcontracted<br>☐ Supplied from third party | ☐ Not allowed<br>☒ Manual request<br>☐ Web service<br>☐ Query script | ☒ Annually or less frequently<br>☒ Monthly to annually<br>☒ More frequently than monthly<br>☒ Ad hoc |

*Table 13 : Summary of responses by developers regarding well testing exploration data.*

**Models**

All the developers manage conceptual models and 92% of the developers manage hydrogeological models and geothermal resource assessments. Between 50 and 60% of the developers collect and/or manage geomechanical and coupled models.

In most cases, models are managed in-house although few developers rely on subcontractors or supply from third party (*Table 14*). It is interesting that management of numerical models, geomechanical models and coupled models appears to be more frequently subcontracted than management of other models or other data.

In most cases (80-100%), developers use commercial modelling software for modelling, except for for conceptual modelling and resource assessment (60%). Although use of commercial software is the most common practise, several developers also use either proprietary or open source software for the modelling.

Developers commonly store models in directories on servers. However, other methods are also commonly used (e.g. local directory, local database, database on server).

Developers generally define all their models as confidential data. However, about 75% of the developers allow access to their models upon manual request.

| Data | Sofware | Storage | Mandatory submission | Confidentiality | Data Management | Access | Frequency |
|---|---|---|---|---|---|---|---|
| Models | ☒ Proprietary<br>☒ Commercial<br>☒ Open source<br>☐ Other: | ☒ Local Directory<br>☒ Directory on server<br>☒ Local database<br>☒ Database on server<br>☒ Open source | ☐ Yes<br>☒ No | ☐ Public<br>☒ Confidential<br>☐ Time-bound | ☒ In house<br>☒ Subcontracted<br>☐ Supplied from third party | ☒ Not allowed<br>☒ Manual request<br>☐ Web service<br>☐ Query script | ☒ Annually or less frequently<br>☐ Monthly to annually<br>☐ More frequently than monthly<br>☒ Ad hoc |

*Table 14 : Summary of responses by* developers *regarding models.*

System architecture, storage, quality control and ownership

Majority of developers have an IS in place (84%) and the systems are generally being managed and coordinated by the developers themselves. The same initial motivations are listed for developers as for the Regulators, both include resource assessment, use and data share as well as the management, security and consistency of datasets. Storage of data is the main function of the IS in place.

For nearly 70% of developers the IS are setup for unstructured data. Unstructured data is not less valuable or useable data than structured data, but it will require more effort to process with other data. The data management systems used by developers therefore need to facilitate diverse data. To some extent, there seems to be limited effort to construct a relational database for exploration data and it is possibly technically and economically challenging to have a database management system to support so diverse data formats.

ESRI – Geodatabase and Microsoft SQL server are used for 50% of the surveyed companies, other types of database management systems used are: Microsoft Access, Oracle, My SQL, PostgreSQL, GD Manager. The data and databases used by developers are generally (86%) hosted locally. An important aspect of geothermal exploration datasets is confidentiality and processing and interpretation of the data is a key factor in planning and development of geothermal exploration fields. Developers do not seem to be using semantical standards to support the IS management. Nevertheless, for almost half of them, the management of the IS in place is a part of a quality management system, ISO 9001:2015.

Most developers use discipline experts to control the quality of the data (87%) but there is still a large part of the data that is not geospatially coherent (43%). This can affect the use of data and specially the ability to integrate data and allow for interdisciplinary interpretation.

Accessibility and dissemination

For 3/4 of the companies, the data are not accessible from entities outside the company. When accessible from outside the company, the access constraints to the IS are restricted.

To visualize and share data, very different types of software (in house, open source, third party) and capabilities (e.g. share and visualize, share + visualize + process) are used.

Cost of Data information system installation

Three different types of Information Systems seem to co-exist within private companies and display different investment levels:

1. “Light” IS developed in less than a year with limited initial investment (less than $100,000) and staff (less 25person months). Associated operational costs are in the $10,000 - $100,000 per year range and involves less than three person per year.
2. “Moderate” IS developed in one to five years with initial investments in the $100,000 to $1,000,000 range and between 26 and 50 person months. Associated operational costs are in the $10,000 - $100,000 per year range and involves four to nine persons per year.
3. “Heavy” IS developed in more five years with initial investments in the $100,000 to $1,000,000 range and more than 100 person months. Associated operational costs are in the $10,000 - $100,000 per year range and involves four to nine persons per year.

Reasons for the cost differences include types/amount of data dealt with, company policies for data storage/sharing and IT infrastructure, integration/dependence on other ISs.

## 3 RECOMMENDATIONS FOR BEST PRACTICES ON GEOTHERMAL EXPLORATION DATA MANAGEMENT

The output of this work is a suggestion of industry best practices and benchmarks to be used to define the scope of future technical assistance and capacity building efforts on geothermal data management to accompany and support World Bank projects involving geothermal exploration.

### 3.1 Regulators

In order to develop an IS for managing geothermal exploration data for Regulators, we suggest the following best practices:

Regulatory Framework:

- Promote, develop or improve national/regional regulatory framework in compliance with the geothermal market to harmonize practices of the different players and hence further develop the geothermal energy utilization.
- Develop or improve national/regional standardized workflow to publish exploration data, in agreement with laws and regulations. Standardized workflow helps data collection and reporting but also management of data, both for developers and regulatory entities.
- Develop or improve cooperation between regulatory entities and developers to inform about and facilitate the compliance of developers with submission of exploration data, and minimize enforcement.
- Favour public release of exploration data in the regulatory framework to attract investors and developers.

Data Types:

- Geological data: Structural data, Lithological data, Hydrogeological data, Geothermal manifestations (hydrothermal alteration, springs, etc.), Geological maps, Geological cross sections, Remote sensing data
- Geophysical data: Gravimetric surveys, Geomagnetic surveys, Electromagnetic surveys, Passive seismic surveys, Active seismic surveys, Ground temperature/heat flow mapping
- Geochemical data: Location, temperature and type of active geothermal manifestations, Fluid/gas composition, Fluid/gas samples
- Borehole data: Well design (trajectory, diameter, casing scheme), Drilling parameters, Drilling reports, Well completion reports, Descriptions of cores/cuttings
- Well logging data: Caliper logs, Cement bound logs, Temperature and pressure logs, Lithology logs, Well testing data, Fluid temperature, Pressure/drawdown, Well discharge (steam and liquid), Fluid movement (e.g. spinner logs), Tracer flow testing, Well test report
- Models: Conceptual models, Geophysical models, Hydrogeological models, Thermal models, Geothermal resource assessments

Data Format:

- Data collected and managed in raw, processed and interpreted formats in order to allow for reprocessing, visualization and calculations.
- Data stored in an international or industry-wide standards, if they exist. If not, best practice is to develop a national standard (e.g. USGIN standard as developed in the US).
- Data format must be digital as much as practically possible (e.g. photos of cores, scans of well logs, scans of paper reports) or else metadata of the actual data must be provided.
- Specify the data standards and references (e.g. INSPIRE in Europe, NGDS in the US), especially commonly agreed terminologies to semantically harmonize existing databases. Interoperable formats help in order to make data transmission process automatic and timesaving.
- Data collected on an annual basis in full compliance to regulatory framework. Failure to do so could potentially put a risk the effectiveness of the whole data management system.

System Architecture, Storage, QC and Ownership

- Develop a central system that produces new/restructures information i.e. with both a harvesting and diffusion systems.
- Where geothermal developers have developed a reliable IS managing data of interest, Regulator can develop an IS based on a simple feature/index approach (e.g. metadata like the NGDS system in the US). It has the advantage to minimize the investment in the harvesting system and to focus investments on data diffusion but also the main drawback to heavily rely on the secure storage and availability of data through third parties.
- Store data on a fully owned IS. If that requires to develop a new data management system, best practice is to develop relational databases with existing commercial software (e.g. ESRI Geodatabase, Microsoft SQL server, Oracle) or open source solutions (e.g. RDBMS PostGreSQL).
- The quality control (QC) of data is an important issue for the support and management of large geothermal exploration datasets. Issues concerning data cleansing, restrict input, completeness, validity, accuracy and consistency are important and standardized methods for QC must be developed.
- Ownership of data can be sensitive regarding to competitive issues. Clear regulatory framework for the data collected and stored by regulatory entities is very important. The IS must be able to handle different data ownership levels in order to comply with laws and regulations.

Accessibility and dissemination

- The dissemination part of the IS provides access to the data. Recommendation is to have data available 24/7 and to avoid manual de-archiving procedures. For this purpose, recommendation is to have data as FAIR (Findable, Accessible, Interoperable, Reusable) as practically possible.

- Recommended to access data through a web service designed in-house or by a third-party. However, a most-cost effective, but also risky option, is to use an open source platform/software. The main advantage of open source option is their free installation, but also the lack of maintenance fees for their update. Another advantage is the speed of evolution and bug fixes, when the community is very active.

Cost of installation and maintenance

- An IS must be planned to be operational for more than 20 years and take between 1 and 5 years to develop as proposed.
- Estimated initial investment (including salaries, cost for hardware and software) is 100,000 - 1,000,000 USD involving 26-50 person months (cumulatively since the commissioning of the System).
- Estimated budget the maintenance and operation is 10,000 – 100,000 USD/year, involving less than 3 persons/year.
- Estimated budget for new developments of the system is $10,000 - $100,000/year.
- If an IS based on a Simple Feature / Index approach is developed, costs are likely to be lower as the harvesting system and associated QC is kept to the minimum.

### 3.2 Developers

In order to develop an IS for managing geothermal exploration data for developers, we suggest the following best practices:

Data Types and Format

- Developers gather data to assess geothermal potential and facilitate use of geothermal/subsurface data as their initial motivation. Only about a quarter of the developers mentioned legal requirements as their initial motivation for implementation of IS for exploration data. The responds to the questionnaire shows that the following data are routinely collected in mature geothermal markets and is commonly shared with Regulators.
- Best practice is to collect such data in raw, processed and interpreted formats in order to allow for reprocessing, visualization and calculations. Whatever the format of the data, the best practice is to store the data in an industry-wide standards, if they exist. If not, best practice is to adopt an international or national standard, if provided by the Regulator.
- Data format shall be digital as much as practically possible (e.g. photos of cores, scans of well logs, scans of paper reports). If not, metadata of the actual data shall be provided.
- Data collection procedures shall be clearly defined as part of the company data quality management system. It shall specify the data model to be used (e.g. type, format, terminology). Regular engagements with data providers (internal and/or external, if outsourced) shall be scheduled to ensure full compliance. Best practice is to collect such data on an annual basis.
- The development of both a harvesting and diffusion systems is necessary. All data are collected from the data provider, quality controlled and stored into the harvesting system. It is subsequently pushed into a diffusion system in charge of generating new content and used for the dissemination.

System architecture, Storage, quality control and ownership

- Best practice is to develop an IS with a central system that produces new / restructures information i.e. with both a harvesting and diffusion systems.
- An alternative is to develop an IS based on a simple feature / index approach to minimize the investment in the harvesting system and to focus investments on data diffusion but also the main drawback to heavily rely on the secure storage and availability of data through the selected third parties.
- Store data on a fully owned IS using relational databases with existing commercial software. The industry does not seem to prefer any particular type of software developer (e.g. ESRI Geodatabase, Microsoft SQL server, Oracle, RDBMS PostGreSQL) and choice has to be made at the company level.
- The quality control (QC) of data concerning data cleansing, restrict input, completeness, validity, accuracy and consistency are important and standardized methods that must be developed and performed by the developer or a third party consultant at the data collection stage.
- Data remains the property of the developer.

Accessibility and dissemination

- Recommended is that data accessibility is managed and controlled by the developer, data is available 24/7 and manual de-archiving procedures are avoided. Depending on the company policy and regulatory framework, data or associated metadata may be accessible from entities outside the company. For this purpose, recommendation is to have data as FAIR (Findable, Accessible, Interoperable, Reusable) as practically possible.
- Access to data is through a web service designed in-house or by a third-party. However, a most-cost effective but also risky option is to use an open source platform/software. The main advantage of open source option is their free installation, but also the lack of maintenance fees for their update. Another advantage is the speed of evolution and bug fixes, when the community is very active

Cost of installation and maintenance

- An IS must be planned to be operational for more than 20 years and take between 1 and 5 years to develop as proposed.
- Estimated initial investment (including salaries, cost for hardware and software) is $100,000$ - $1,000,000, involving more than 100 person months (cumulatively since the commissioning of the System).
- Estimated budget for its maintenance and operation (including salaries, cost for hardware and software) is $10,000 – $100,000/year, involving less than 5-9 persons/year.
- Estimated budget for new developments of the system is $10,000 - $100,000/year.

## CONCLUSION

Setting up of an Information System (IS) is the best practice to systematically and securely manage exploration data. Recommendations with respect to the regulatory framework, data types, data collection methodologies, data storage, data quality control, data accessibility and dissemination, IS architecture, financial investments and human resources have been proposed to develop a state-of-the art IS. These findings will guide the design of future technical assistance programs for beneficiaries of World Bank support to geothermal exploration activities. It is our belief that the findings will also be beneficial for the geothermal sector at large.

## ACKNOWLEDGEMENTS

The authors wish to thank all the developers and Regulators that participated to the survey on their practices for geothermal data management. We also would like to thank Alexander Richter, president of the International Geothermal Association for for his support throughout the project.